\documentstyle[prl,aps,epsfig,floats]{revtex} 
\title{\bf A note on bound entanglement and local realism.} 
\author{Dagomir Kaszlikowski$^{1,2}$, Marek \.Zukowski$^{3}$, Piotr Gnaci\'nski$^{3}$.} 
\address{$^1$ Department of Physics, National University of Singapore, 10 Kent Ridge Crescent, 
Singapore 119260, $^2$ Instytut Fizyki Do\'swiadczalnej, 
Uniwersytet Gda\'nski, PL-80-952, Gda\'nsk, Poland,  
$^3$ Instytut Fizyki Teoretycznej i Astrofizyki, 
Uniwersytet Gda\'nski, PL-80-952, Gda\'nsk, Poland.} 
\begin{document} 
\maketitle  {\bf Abstract}: We show using a numerical approach  that gives {\it necessary and sufficient} conditions for the existence of local realism, 
that the bound entangled state presented in Bennett et. al. Phys. Rev. Lett. {\bf 82}, 5385 (1999) 
admits a local and realistic description. We also find the lowest possible amount of some 
appropriate entangled state that must be ad-mixed to the bound entangled state so that 
the resulting density operator has no local and realistic description and as such can 
be useful in quantum communication and quantum computation.  
\section{Introduction} 
Since the discovery of the protocol of quantum teleportation in 1993 \cite{BBC93} it 
has been shown that quantum entanglement is a fundamental resource in quantum communication. 
Some recent research suggests that the phenomenon of quantum entanglement can be also viewed 
as a resource in quantum computation \cite{COMPUTATION}.  

The best source of quantum entanglement 
is, of course, a pure maximally entangled state. However, if one thinks about quantum communication 
and quantum computation in a noisy environment than it is necessary to deal with mixed states. It 
has been shown that for a wide class of mixed entangled states by distillation protocol \cite{DISTILLATION} 
one can extract useful entanglement that can be then used as a resource in quantum communication 
and quantum computation. Nevertheless, there still exists a set of entangled mixed states 
which cannot be distilled \cite{BOUND} . This phenomenon has been called bound entanglement. 
Although, these states are undistillable they are not completely useless in quantum communication. 
In \cite{ACTIVATION} it has been shown that some family of bound entangled states can be useful 
in conclusive quantum teleportation process.  

The discovery of bound entangled states has raised 
a question whether they can be described in a local and realistic way, i.e., if there exists a 
Bell-type experiment in which predictions of local realism are violated by quantum mechanical ones. 
It has been Peres who first conjectured that such states should admit a local and realistic 
description \cite{PERES}. However, as there are no known Bell inequalities being a sufficient 
and necessary condition for the existence of local realism for systems described by the tensor 
product of two Hilbert spaces each of the dimension greater than two (we know that lowest dimensional 
bound entangled state lives in a tensor product of two Hilbert spaces of the dimension equal to three), 
it could not be proved\footnote{Recently discovered Bell inequalities for higher dimensional quantum 
systems \cite{CHQUTRITS,COLLINS} have not been proved to be sufficient condition for the existence of 
local realism}. Therefore,  the question has remained open.  

In this paper we show that using numerical 
method of linear optimization we can answer the above question. Although, here we investigate the specific 
bound entangled state the presented method can be successfully applied to any bound entangled state 
(in general, to any mixed state). The strength of the method lies in the fact that it gives {\it necessary 
and sufficient} conditions for the existence of local and realistic description of the investigated quantum 
system.  

Additionally we calculate how much of some optimally defined entangled state must be ad-mixed to 
the investigated bound entangled state so that the created mixture does not have a local and realistic 
description anymore.  
\section{Construction of the bound entangled state} 
In \cite{BENNETT} it has been shown a simple and elegant way to construct 
bound entangled states. The idea is based on the theorem proved by Horodecki 
\cite{HORODECKI} and on the notion of unextendible product basis (UPB) \cite{BENNETT}.  

In short, the Horodecki theorem states that if a density matrix $\rho$ of two quantum systems 
$A$ and $B$ is separable then there exists a family of product vectors, 
say $|\phi_{k}\rangle|\psi_{k}\rangle$ (first ket refers to system $A$ and the second one to 
system $B$), spanning the range of $\rho$ and such that vectors $|\phi_{k}\rangle|\psi_{k}\rangle^{*}$ 
span the range of partially transposed matrix $\rho^{T_B}$, where $T_B$ means the transposition with 
respect to the subsystem B and the star denotes complex conjugation.  

An unextendible product basis 
is a set of $d$ orthogonal product vectors belonging to $N\times M$ dimensional Hilbert space, 
where $d<N\times M$, and of the property that any orthogonal vector to them must be entangled.  

Using the definition of UPB and the Horodecki theorem one can easily construct bound entangled states. 
Let us assume that we have a Hilbert space being a tensor product of two $N$ dimensional Hilbert spaces. 
In this space we find some unextendible product basis consisting of $d$ product 
vectors $|\phi_{k}\rangle|\psi_{k}\rangle$, where $k=1,2,\dots,d$. Denoting by $\hat{P}$ the 
projector on the subspace spanned by them one can define the following density matrix $\rho$ 
\begin{eqnarray} 
&\rho={1\over N^2-d}(\hat{I}-\hat{P})&. 
\end{eqnarray} 
The support of this matrix (space spanned by its eigenvectors) lies in a subspace perpendicular 
to the subspace spanned by the vectors from the unextendible product basis, therefore it is 
spanned by $N^2-d$ entangled orthonormal vectors. The Horodecki theorem says that this state must 
be entangled and as it can be directly checked its partial transposition is a positive operator. 
Hence we have obtained a bound entangled density matrix.  

In this paper we investigate a particular bound entangled state consisting 
of two subsystems each living in a three dimensional Hilbert space. As the UPB we choose the following 
set of orthonormal vectors \cite{BENNETT}  
\begin{eqnarray} 
&&|v_0\rangle = {1\over\sqrt 2}(|0\rangle|0\rangle- |0\rangle|1\rangle)\nonumber\\ 
&&|v_1\rangle = {1\over\sqrt 2}(|0\rangle |2\rangle - |1\rangle|2\rangle)\nonumber\\ 
&&|v_2\rangle = {1\over\sqrt 2}(|2\rangle|1\rangle - |2\rangle|2\rangle)\nonumber\\ 
&&|v_3\rangle = ({1\over\sqrt 2}(|1\rangle|0\rangle- |2\rangle|0\rangle)\nonumber\\ 
&&|v_4\rangle = {1\over\sqrt 2}(|0\rangle+|1\rangle+|2\rangle) (|0\rangle+|1\rangle+|2\rangle, 
\label{UPB1} 
\end{eqnarray} 
and follow the recipe described above. The resulting density matrix can be written as follows  
\begin{eqnarray} 
\rho_{B}={1\over 4}(|v_5\rangle\langle v_5|+|v_6\rangle\langle v_6|+|v_7\rangle\langle v_7|+ |v_8\rangle\langle v_8|), 
\end{eqnarray} 
where  
\begin{eqnarray} 
&&|v_5\rangle = {1\over\sqrt 2}(|v_{0}^{+}\rangle-|v_{1}^{+}\rangle)\nonumber\\ 
&&|v_6\rangle = {1\over\sqrt 2}(|v_{2}^{+}\rangle-|v_{3}^{+}\rangle)\nonumber\\ 
&&|v_7\rangle = {1\over 2}(|v_{0}^{+}\rangle+ |v_{1}^{+}\rangle - |v_{2}^{+}\rangle - |v_{3}^{+}\rangle)\nonumber\\ 
&&|v_8\rangle ={1\over 6}(|v_{0}^{+}\rangle+ |v_{1}^{+}\rangle + |v_{2}^{+}\rangle + |v_{3}^{+}\rangle) - {2\sqrt2\over 3}|1\rangle|1\rangle, 
\label{UPB2} 
\end{eqnarray} 
and where vectors $|v_{k}^{+}\rangle$ ($k=0,1,\dots, 3$) are made of the states belonging to (\ref{UPB1}) 
by changing the sign $-$ to $+$. For instance, $|v_{0}^{+}\rangle ={1\over\sqrt 2}(|0\rangle|0\rangle + 
|0\rangle|1\rangle$ and so on.  

The above vectors are normalized, orthogonal to each other and orthogonal 
to vectors forming the UPB basis, i.e., vectors $|v_l\rangle$ with $l=0,1,2,\dots, 4$. Therefore, they span 
the orthogonal complement of the subspace associated with the UPB basis. As a whole, the vectors $|v_k\rangle$ 
with $k=0,1,\dots, 8$ form an orthonormal basis in Hilbert space.  
\section{Local Hidden Variables}  
On the 
above mixed state $\rho_{B}$ one can perform a Bell type experiment in which two spatially 
separated observers Alice and Bob measure some trichotomic observables. In this paper we consider the 
case in which both observers are allowed to measure only two noncommuting trichotomic observables, 
which we denote by $\hat{A}_1, \hat{A}_2$ for Alice and by $\hat{B}_1, \hat{B}_2$ for Bob.  

Any 
trichotomic observable in a three dimensional Hilbert space can be obtained by a rotation of some orthogonal 
basis by means of a unitary transformation belonging to $SU(3)$ group. $SU(3)$ group is a set of unitary matrices 
with the determinant equal to one and which depend on eight real parameters 
$\phi_1,\phi_2,\dots,\phi_8$, which we will denote as a vector $\vec{\phi}=(\phi_1,\phi_2,\dots,\phi_8)$. Suppose
that the resolution of unity at Alice's side consists of three orthogonal projectors 
$\hat{P}_k=|k\rangle\langle k|$ and that at Bob's side of projectors $\hat{Q}_l=|l\rangle\langle l|$ 
($k,l=1,2,3$). Then arbitrary trichotomic observables for Alice and Bob read  
\begin{eqnarray} 
&&\hat{A}_i=a_{1}^{i}U(\vec{\phi}_{A}^{i})\hat{P}_1U(\vec{\phi}_{A}^{i})^{\dagger} + 
a_{2}^{i}U(\vec{\phi}_{A}^{i})\hat{P}_2U(\vec{\phi}_{A}^{i})^{\dagger} + 
a_{3}^{i}U(\vec{\phi}_{A}^{i})\hat{P}_3U(\vec{\phi}_{A}^{i})^{\dagger}\nonumber\\ 
&&\hat{B}_j=b_{1}^{j}V(\vec{\phi}_{B}^{j})\hat{P}_1V(\vec{\phi}_{B}^{j})^{\dagger} + 
b_{2}^{j}V(\vec{\phi}_{B}^{j})\hat{P}_2V(\vec{\phi}_{B}^{j})^{\dagger} + 
b_{3}^{j}V(\vec{\phi}_{B}^{j})\hat{P}_3V(\vec{\phi}_{B}^{j})^{\dagger}, 
\label{observables} 
\end{eqnarray} 
where $U(\vec{\phi}_{A}^{i}),V(\vec{\phi}_{B}^{j})$ $(i,j=1,2)$ are 
members of $SU(3)$ group and where the numbers $a_{k}^{i},b_{k}^{j}$ ($k=1,2,3$) 
are the eigenvalues of appropriate observables. Of course each observable $A_i$ and $B_j$ 
depends on vectors $\vec{\phi}_{A}^{i}$ and $\vec{\phi}_{B}^{j}$ but to shorten 
notation we will not write it explicitly.  

The probability $P_{QM}(a_{k}^{i},b_{l}^{j}|\vec{\phi}_{A}^{i},
\vec{\phi}_{B}^{j})$ of obtaining the pair of eigenvalues $a_{k}^{i},b_{l}^{j}$ while measuring the 
observables $\hat{A}_{i}, \hat{B}_{j}$ on the density matrix $\rho$ can be calculated in a standard way as  
\begin{eqnarray} 
&&P_{QM}(a_{k}^{i},b_{l}^{j}|\vec{\phi}_{A}^{i},\vec{\phi}_{B}^{j})= 
Tr(U(\vec{\phi}_{A}^{i})\hat{P}_kU(\vec{\phi}_{A}^{i})^{\dagger}\otimes 
V(\vec{\phi}_{B}^{j})\hat{Q}_lV(\vec{\phi}_{B}^{j})^{\dagger}\rho). 
\label{set} 
\end{eqnarray} 
Obviously, there are 36 such probabilities.  

A local and realistic description of the presented 
quantum experiment is equivalent to the existence of a joint probability distribution, which returns 
all 36 quantum probabilities as marginals \cite{FINE}. Let us denote this hypothetical probability 
distribution as $P_{LR}(a_{k}^{1},a_{l}^{2},b_{m}^{1},b_{n}^{2})$, where $k,l,m,n=1,2,3$. It consists of 
81 non-negative numbers summing up to one. The marginals are given by the following set of equations  
\begin{eqnarray} 
&&P_{LR}(a_{k}^{1},b_{m}^{1})=\sum_{l}\sum_{n}P_{LR}(a_{k}^{1},a_{l}^{2},b_{m}^{1},b_{n}^{2})\nonumber\\ 
&&P_{LR}(a_{k}^{1},b_{n}^{2})=\sum_{l}\sum_{m}P_{LR}(a_{k}^{1},a_{l}^{2},b_{m}^{1},b_{n}^{2})\nonumber\\ 
&&P_{LR}(a_{l}^{2},b_{m}^{1})=\sum_{k}\sum_{n}P_{LR}(a_{k}^{1},a_{l}^{2},b_{m}^{1},b_{n}^{2})\nonumber\\ 
&&P_{LR}(a_{l}^{2},b_{n}^{2})=\sum_{k}\sum_{m}P_{LR}(a_{k}^{1},a_{l}^{2},b_{m}^{1},b_{n}^{2}). 
\label{constraints} 
\end{eqnarray} 
A quantum experiment admits a local and realistic description {\it if and only if} the above 
marginals can be made equal to quantum ones \cite{FINE}, i.e., $P_{LR}(a_{k}^{i},b_{l}^{j})=
P_{QM}(a_{k}^{i},b_{l}^{j}|\vec{\phi}_{A}^{i},\vec{\phi}_{B}^{j})$.  

As there are not known Bell inequalities 
for two three dimensional quantum systems that are necessary and sufficient condition for the existence of 
local realism one is forced to resort to numerical methods. The problem can be solved numerically by means of 
{\it linear programming} method \cite{ZUKOWSKI,KASZLIKOWSKI}. Linear programming is a method of finding the 
maximum of a linear multi-variable function, which domain is a convex set. It relies on the fact that the 
maximum is reached (if it exists) in one of the vertices of the domain. Moreover, linear programming can be 
used (by setting the function being maximized to be constant) to check if the set of linear equations that 
form the boundaries of a convex set (domain) has a solution. In this case the lack of the solution means that 
the considered convex set is empty, i.e., it has no "interior".  

Therefore, linear programming can be successfully 
applied to our case. Indeed, we are interested in finding a local realistic probability distribution 
$P_{LR}(a_{k}^{1},a_{l}^{2},b_{m}^{1},b_{n}^{2})$, i.e., 81 non-negative numbers, returning quantum 
probabilities as marginals. The set of equations (\ref{constraints}) plus the condition that the probabilities 
sum up to one defines, for the given set of observables, a convex set in an 82 dimensional real vector space. 
Now, applying linear programming procedure we can check if the set is empty. If not, then there exists a local 
and realistic description of the experiment.  

Thus, for each choice of the observables 
$\hat{A}_1,\hat{A}_2, \hat{B}_1,\hat{B}_2$ we apply the numerical linear programming procedure. 
In our calculations we have used the state-of-the-art HOPDM 2.30 procedure \cite{GONDZIO}. The quadruples 
of observables (two for Alice and two for Bob) have been chosen randomly $10^6$ times. Local and realistic 
description has existed for all these cases. Therefore, it suggests that the entanglement contained in this 
state is too weak to violate local realism and as such it is a strong evidence supporting Peres' conjecture 
\cite{PERES} that bound entangled states admits a local and realistic description.   
\section{Extracting entanglement}  
Although the number ($10^6$) of Bell experiments simulated numerically is huge it may be that one can 
still find some quadruple of observables for which there is a violation of local realism by the bound entangled 
state $\rho_{B}$. Therefore, it is desirable to  support the above result using some additional calculations. 
To this end we propose  the following procedure. 

We ask what is the least amount $F$ ($0\leq F\leq 1$) of some  
entangled states $|\psi\rangle\langle\psi|$ that has to be admixed to the state $\rho_{B}$  
so that the resulting state $\rho(F)=(1-F)\rho_{B}+F|\psi\rangle\langle\psi|$ has not a local realistic 
description anymore. It is clear that in such a case our  
numerical procedure for large $F$ cannot return a local hidden variable  model for any set of local settings.  
However, for lower $F$'s, as it turns out, it does find such models. Since numerical linear optimization is 
a highly reliable procedure, and we additionally use another procedure (see below) to find  the optimal 
parameters of the problem for the existence of local hidden variable models, the results obtained in this  
way are much more definitive than the "lottery" approach presented above.  Also they in a way measure 
the "robustness" of the local realistic models for the bound entangled state.    

As the support of the state $\rho_{B}$ is spanned by the vectors $|v_i\rangle$ ($i=5,6,7,8$) 
it is naturally to consider the state $|\psi\rangle$ as a superposition of these vectors, i.e., 
$|\psi\rangle = \sum_{i=5}^{8}a_i|v_{i}\rangle$, where $\sum_{i=5}^{8}|a_i|^2=1$. It is convenient 
to parametrize the complex numbers $a_i$ by six angles $\psi,\theta,\phi,\chi_1,\chi_2,\chi_3$ in the 
following way: $a_5=\sin\psi\sin\theta\cos\phi, a_6=\exp(i\chi_1)\sin\psi\sin\theta\sin\phi, 
a_7=\exp(i\chi_2)\sin\psi\cos\theta, a_8=\exp(i\chi_3)\cos\psi$.  

To find the optimal state $|\psi\rangle$ (optimal in the sense defined above) for every 
choice of angles $\psi,\theta,\phi,\chi_1,\chi_2,\chi_3$ and the observables 
$A_1,A_2,B_1,B_2$ (we remind that each observable depends on 8 angles) we calculate using linear 
programming procedure HOPDM 2.30 the maximal value of the parameter $F$,  
which now depends on 38 angles (32 angles defining observables and 6 angles defining 
the state $|\psi\rangle$), below which there exists a local and realistic description of 
the experiment. 
This way we obtain the 38 variable function which minimum $F_{min}$ can be found by the 
so called amoeba procedure utilising downhil simplex method \cite{MELDER}. Although the amoeba
is a good minimization procedure there is no guarantee that the found minimum is a global one
(a problem encountered in any numerical minimization). To reduce the risk of finding the local
minimum the procedure has been run many times with randomly chosen initial conditions. 

Calculations show that the minimal possible $F$ equals $F_{min}=0.509651$. This occurs 
for the state  $|\psi\rangle = {1\over\sqrt 2}(|v_5\rangle + |v_6\rangle)$  
(the analytical form of the state $|\psi\rangle$ has been obtained using  
the numerical results and then, for  additional confirmation, the numerical  
calculations have been performed again for  the guessed state without the minimization over 
the angles characterizing it yielding the same $F_{min}$).   

It can be directly checked that the state $\rho(F)$ for any $F\neq 0$ is negative under partial 
transposition. 
Therefore, we have the situation in which the family of mixed states $\rho(F)$ for $0\leq F < F_{min}$ is entangled 
but has a  local realistic description. This once again shows that non-separability is not equivalent to 
lack of classical description.   

It is instructive to calculate the degree of entanglement of the state 
$|\psi\rangle$ defined as the ${3\over 2}(1-Tr_{B}( (Tr_{A}(|\psi\rangle\langle \psi|))^2))$, where for 
instance $Tr_{A}$ denotes the trace with respect to Hilbert space of Alice's subsystem. Trivial algebra 
gives us ${15\over 16}$. It can be checked  numerically that this is the most entangled state that can be  
obtained by superposing the states  $|v_{i}\rangle$ ($i=5,6,7,8$). This is an interesting confirmation of 
something that one could intuitively expect. The most  efficient way to  get to a region in which  there is 
no  local realistic description is via  an admixture of the most entangled pure state that lives in the subspace 
of  the Hilbert space which is  used to construct the considered  bound entangled state.      
\section{Conclusions}  
The state $\rho_B$ cannot be distilled, which means that the entanglement is hidden in the state 
too deep to be recovered by local quantum operations. Moreover, the presented here numerical computations 
strongly support the hypothesis that the bound entangled state $\rho_{B}$ is describable by local hidden 
variables or in other words that the correlations observed in a Bell-type experiment with this state can be 
simulated classically. The large value of  $F$, the admixutre of optimal entangled state required to get violations of local realism, 
clearly indicates that the considered bound state is not even close to a border of the realm of states possessing a local realistic model. 
It is well inside this realm.

It has been recently shown that {\it multipartite} (more than seven quantum systems)  
bound entangled states violate local realism \cite{DUR}. In the view of the results presented here, i.e., 
no violation of local realism by the {\it bipartite} bound entangled state $\rho_{B}$, one sees that one is 
far away from a complete understanding of the structure of bound entanglement. We hope that it will stimulate 
further research in this direction.  
\\
\\
{\it Acknowledgements}  

We wish to thank prof. Bennett for suggestions concerning the choice of the bound entangled state. 
DK would like to thank prof. Bennett for  
stimulating discussion during the workshop in Singapore and Hazlinda Nuron for 
support.  MZ and DK are supported by KBN grant No. 5 P03B 088 20.   
    
\end{document}